\documentclass[reprint,amsmath,amssymb,aps,prc,superscriptaddress]{revtex4-2}

\usepackage{graphicx}
\usepackage{dcolumn}
\usepackage{bm}
\usepackage[colorlinks=true,
    allcolors=BlueViolet,
    ]%
    {hyperref}

\newcommand{\amend}[1]{\textcolor{black}{#1}}

\usepackage{multirow}
\usepackage[dvipsnames]{xcolor}
\usepackage{soul}
\sethlcolor{white}
\usepackage{tabularx}

\newcommand{\K}{{$^{40}$K}}

\newcommand{\Ca}{{$^{40}$Ca}}
\newcommand{\Ar}{{$^{40}$Ar}}

\newcommand{\ArAr}{\mbox{$^{40}$Ar/$^{39}$Ar}}
\newcommand{\KAr}{\mbox{K/Ar}}

\newcommand{\ECStar}{\mbox{EC$^{*}$}}
\newcommand{\EC}{\mbox{EC$^{0}$}}
\newcommand{\PEC}{\mbox{$I_\text{EC$^{0}$}$}}
\newcommand{\PECStar}{\mbox{$I_\text{EC*}$}}
\newcommand{\PBetaMin}{\mbox{$I_{\beta^-}$}}
\newcommand{\PBetaPlus}{\mbox{$I_{\beta^+}$}}
\newcommand{\TMin}{\mbox{$T^-$}}
\newcommand{\TStar}{\mbox{$T^*$}}

\newcommand{\Kalpha}{K$_{\alpha}$}
\newcommand{\Kbeta}{K$_{\beta}$}

\newcommand{\mus}[1]{#1~$\mu$s}

\newcommand{\IECBranchingPer}{0.098}
\newcommand{\IECBranchingStatPer}{0.023}
\newcommand{\IECBranchingSystPer}{0.010}

\newcommand{\IECStarBranchingPer}{10.31}
\newcommand{\IECStarBranchingErrPer}{0.04}
\newcommand{\IBetaMinBranchingPer}{89.59}
\newcommand{\IBetaMinBranchingErrPer}{0.05}
\newcommand{\IBetaPlusExptBranchingPer}{0.00100}
\newcommand{\IBetaPlusExptBranchingErrPer}{0.00013}
\newcommand{\IBetaPlusTheoryBranchingPer}{0.00045}
\newcommand{\IBetaPlusTheoryBranchingErrPer}{0.00012}
\newcommand{\ThalfTotGa}{1.266}
\newcommand{\ThalfTotErrGa}{0.004}

\newcommand{\ratio}{0.0095}
\newcommand{\ratiostat}{0.0022}
\newcommand{\ratiosys}{0.0010}%

\newcommand{\keV}[1]{$#1$~keV}
\newcommand{\MeV}[1]{$#1$~MeV}

\newcommand{\rhoRes}{\mbox{$ \PEC / \PECStar = \ratio \stackrel{\text{stat}}{\pm} \ratiostat \stackrel{\text{sys}}{\pm} \ratiosys $}}
\newcommand{\PRes}{\mbox{$\PEC = \IECBranchingPer\% \stackrel{\text{stat}}{\pm} \IECBranchingStatPer\% \stackrel{\text{sys}}{\pm} \IECBranchingSystPer\% $}}

\newcommand{\znbb}{$0\nu\beta\beta$}

\newcommand{\AddOakRidgePhys}{Physics Division, Oak Ridge National Laboratory, Oak Ridge, Tennessee 37831, USA}
\newcommand{\AddOakRidgeJINPA}{Joint Institute for Nuclear Physics and Application, Oak Ridge National Laboratory, Oak Ridge, Tennessee 37831, USA}
\newcommand{\AddOakRidgeCNMS}{Center for Nanophase Materials Sciences, Oak Ridge National Laboratory, Oak Ridge, Tennessee 37831, USA}
\newcommand{\AddBerkeleyGeo}{Berkeley Geochronology Center, Berkeley, California 94709, USA}

\begin{document}


\title{Evidence for ground-state electron capture of \texorpdfstring{\K}{40K}}

\date{\today}

\author{L.~Hariasz}
\author{M.~Stukel}
\author{P.C.F.~Di Stefano}\email{distefan@queensu.ca}
\affiliation{Department of Physics, Engineering Physics \& Astronomy, Queen's University, Kingston, Ontario K7L 3N6, Canada}
\author{B.C.~Rasco}
\author{K.P.~Rykaczewski}
\affiliation{\AddOakRidgePhys}
\author{N.T.~Brewer}
\affiliation{\AddOakRidgePhys}
\affiliation{\AddOakRidgeJINPA}
\author{D.W.~Stracener}
\author{Y.~Liu}
\affiliation{\AddOakRidgePhys}
\author{Z.~Gai}
\author{C.~Rouleau}
\affiliation{\AddOakRidgeCNMS}
\author{J.~Carter}
\affiliation{\AddBerkeleyGeo}
\author{J.~Kostensalo}
\affiliation{Natural Resources Institute Finland, Joensuu FI-80100, Finland}
\author{J.~Suhonen}
\affiliation{Department of Physics, University of Jyv\"{a}skyl\"{a}, Jyv\"{a}skyl\"{a} FI-40014, Finland}
\author{H.~Davis}
\author{E.D.~Lukosi}
\affiliation{Department of Nuclear Engineering, University of Tennessee, Knoxville, Tennessee 37996, USA}
\affiliation{Joint Institute for Advanced Materials, University of Tennessee, Knoxville, Tennessee 37996, USA}
\author{K.C.~Goetz}
\affiliation{Nuclear and Extreme Environments Measurement Group, Oak Ridge National Laboratory, Oak Ridge, Tennessee 37831, USA}
\author{R.K.~Grzywacz}
\affiliation{\AddOakRidgePhys}
\affiliation{\AddOakRidgeJINPA}
\affiliation{Department of Physics and Astronomy, University of Tennessee, Knoxville, Tennessee 37996, USA}
\author{M.~Mancuso}
\author{F.~Petricca}
\affiliation{Max-Planck-Institut f\"{u}r Physik, Munich D-80805, Germany}
\author{A.~Fija{\l}kowska}
\affiliation{Faculty of Physics, University of Warsaw, Warsaw PL-02-093, Poland}
\author{M.~Woli{\'n}ska-Cichocka}
\affiliation{\AddOakRidgePhys}
\affiliation{\AddOakRidgeJINPA}
\affiliation{Heavy Ion Laboratory, University of Warsaw, Warsaw PL-02-093, Poland}
\author{J.~Ninkovic}
\author{P.~Lechner}
\affiliation{MPG Semiconductor Laboratory, Munich D-80805, Germany}
\author{R.B.~Ickert}
\affiliation{Department of Earth, Atmospheric, and Planetary Sciences, Purdue University, West Lafayette, Indiana 47907, USA}
\author{L.E.~Morgan}
\affiliation{\amend{U.S. Geological Survey}, Geology, Geophysics, and Geochemistry Science Center, Denver, Colorado 80225, USA}
\author{P.R.~Renne}
\affiliation{\AddBerkeleyGeo}
\affiliation{Department of Earth and Planetary Science, University of California, Berkeley 94720, USA}
\author{I.~Yavin}\email{yavin.itay@gmail.com}

\collaboration{KDK Collaboration}\noaffiliation

\begin{abstract}

Potassium-40 is a widespread, naturally occurring isotope whose radioactivity impacts estimated geological ages spanning billions of years, nuclear structure theory, and subatomic rare-event searches -- including those for dark matter and neutrinoless double-beta decay.
The decays of this long-lived isotope must be precisely known for its use as a geochronometer, and to account for its  presence in  low-background experiments. 
There are several known decay modes for potassium-40, but a predicted electron-capture decay directly to the ground state of argon-40  has never been observed.
The  existence of this decay mode impacts several  fields, while theoretical predictions span an order of magnitude.
Here we report on the first, successful observation of this rare decay mode, obtained by the KDK (potassium decay) Collaboration using a novel combination of a low-threshold X-ray detector surrounded by a  tonne-scale, high-efficiency $\gamma$-ray tagger at Oak Ridge National Laboratory. 
A blinded analysis reveals a distinctly non-zero ratio of intensities of ground-state electron-captures (\PEC) over excited-state ones (\PECStar) of $ \rhoRes $ (68\%CL), with the null hypothesis rejected at 4$\sigma$~[Stukel~\emph{et~al}., \href{https://doi.org/10.1103/PhysRevLett.131.052503}{Phys. Rev. Lett. {\bf 131}, 05203 (2023)}].
In terms of branching ratio, this unambiguous signal yields $\PRes $, roughly half of the commonly used prediction.
This first observation of a third-forbidden unique electron capture improves our understanding of low-energy backgrounds in dark-matter searches and has implications for nuclear-structure calculations. For example, a shell-model based theoretical estimate for the neutrinoless double-beta decay half-life of calcium-48 is increased by a factor of $7^{+3}_{-2}$. Our non-zero measurement shifts geochronological ages by up to a percent; implications are illustrated for Earth and solar system chronologies.

\end{abstract}


\maketitle


\clearpage
\section{Introduction}

\begin{figure}[ht]
  \centering
    \includegraphics[width=\linewidth]{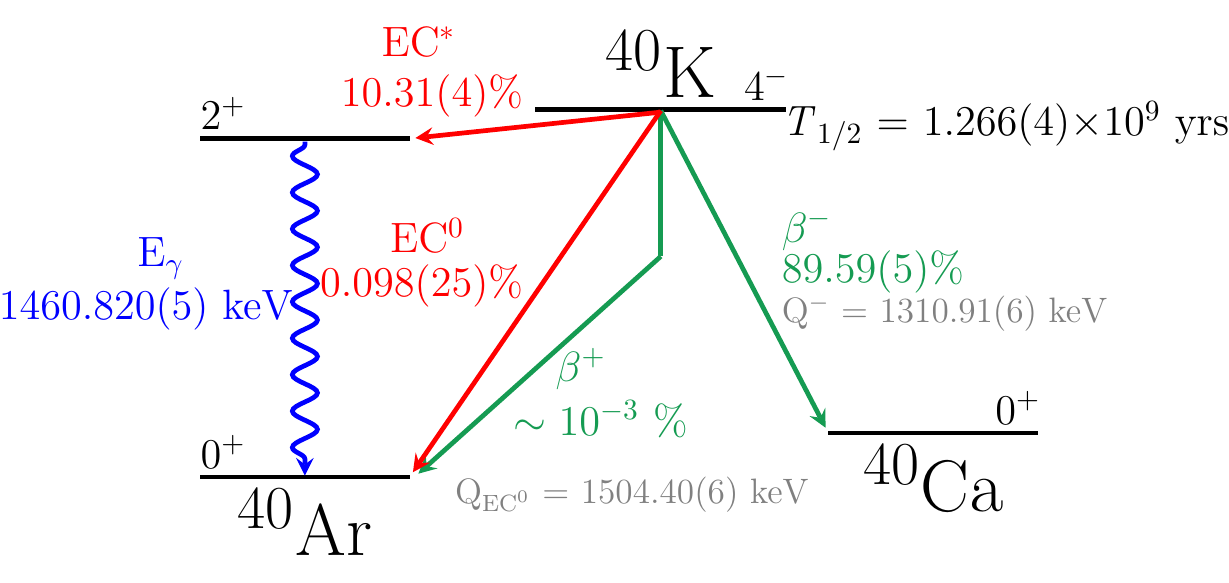}
	
    \caption{\label{Fig:K40_Decay_Scheme_Update_wColor.pdf}\K\ decay scheme, with branching ratios and half-life calculated from our determination of $\PEC / \PECStar$ (this work and~\cite{prl}) and from literature values for $\TMin$ (partial half-life of the $\beta^-$ decay) and $\TStar$ (partial half-life of \ECStar)~\cite{kossert2022activity} (also shown: transition energy~\cite{chen_nuclear_2017}, \amend{Q$_{\text{EC}^0}$ and Q$^{-}$~\cite{wang2021ame}}).}
\end{figure}

Potassium-40 is a common, natural isotope.  It decays mainly by $\beta^-$ decay,  less frequently by electron-capture to an excited state of argon-40 (\ECStar), and very rarely by $\beta^+$ decay (see Fig.~\ref{Fig:K40_Decay_Scheme_Update_wColor.pdf}).  It is a frequent contaminant in various particle detectors,  
and a source of radioactive background in rare-event searches for dark matter~\cite{schumann_direct_2019,pradler_unverified_2013, ANTONELLO20191, adhikari2018background, angloher2022simulation, amare2021annual} and neutrinoless double-beta decay~\cite{EJIRI20191}.  %
Several experiments looking for dark matter are taking draconian steps to purify their NaI detectors of K, and/or deploy vetos to tag the problematic low-energy radiation from \K\ electron-capture which falls in the expected dark matter signal region~\cite{adhikari2018background,ANTONELLO20191,amare2021annual}. This veto method relies on identification of the high-energy $\gamma$-ray from the de-excitation of \Ar\ following an \ECStar\ decay. In addition, the long half-life (\amend{slightly over a billion years}) of this isotope makes it a useful geochronometer via the K/Ar and \ArAr\ dating techniques~\cite{begemann_call_2001,carter2020production,min2000test,renne2010joint}. Finally, the presence of all three modes of $\beta$ decay, some of which are extremely rare third-forbidden unique transitions, make this isotope of particular interest to nuclear structure theory~\cite{EJIRI20191,mougeot_improved_2018,MOUGEOT2019108884}.

Aside from the branches mentioned previously, an electron-capture decay directly to the ground state of \Ar\ (\EC) has also been predicted by some~\cite{engelkemeir_positron_1962,MOUGEOT2019108884,carter2020production,kossert2022activity}, ignored or disputed by others~\cite{min2000test,renne2010joint}, but never observed.  This branch forms a particularly challenging background in rare-event searches as there is no high-energy $\gamma$-ray from de-excitation that could be used to tag the low-energy radiation from the electron capture. Such a background would evade the veto technique mentioned previously.  In addition, this added background has been proposed~\cite{pradler_unverified_2013} as a way to constrain the dark-matter interpretation of the longstanding, but controversial, DAMA/LIBRA Collaboration claim for discovery of dark matter~\cite{bernabei2018first}. From the standpoint of geochronology, the existence of this decay to ground state could mean that samples are up to tens of millions of years older than commonly believed~\cite{carter2020production}.  An empirical frequency of this branch would also inform calculations in nuclear structure theory, including those for neutrinoless double-beta decay half-lives. The KDK (potassium decay) collaboration~\cite{di_stefano_kdk_2017} has carried out the first measurement of \EC~\cite{prl}, using a novel detector configuration~\cite{stukel2021novel}, as detailed in what follows.

\section{Detector and analysis}

\begin{figure*}[ht]
    \centering
    \includegraphics[width=\textwidth]{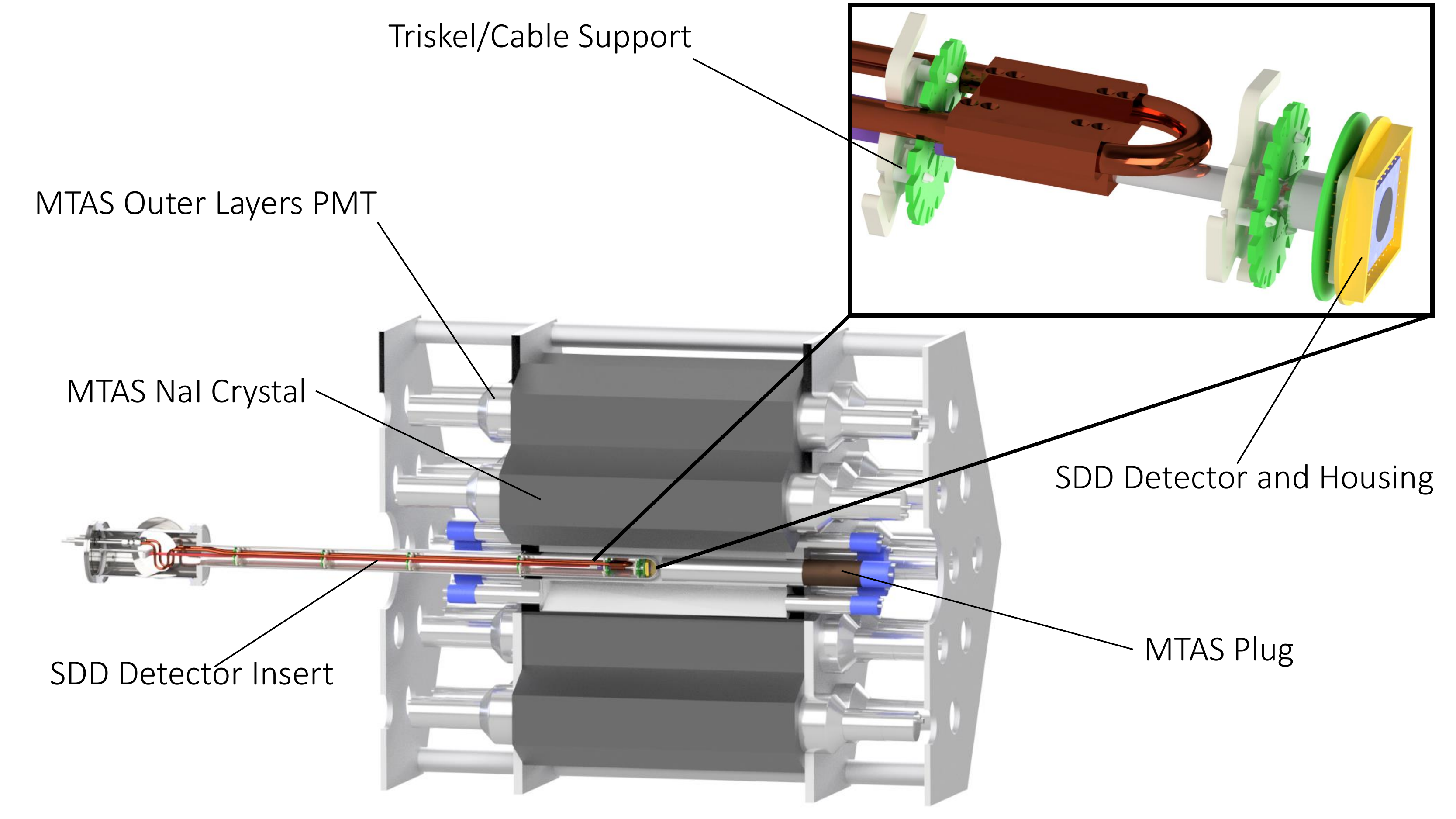}
    
    \caption{\label{fig:MTAS_SDD_schematic}Schematic displaying the cross-section of MTAS, along with the SDD. The SDD housing is centered in MTAS.}
\end{figure*}

\amend{The fully characterized KDK experimental setup~\cite{stukel2021novel} consists of a \K\ source, sensitive X-ray Silicon Drift Detector (SDD) and a near-100\% efficient gamma tagger (the Modular Total Absorption Spectrometer; MTAS~\cite{MTAS_Karny}), as illustrated in Fig.~\ref{fig:MTAS_SDD_schematic}.}

\amend{The enriched \K\ source was made of KCl (16.1(6)$\%$ \K\ abundance in K)  thermally deposited over 1~cm diameter onto a graphite substrate (all uncertainties in this manuscript correspond to a 68\% confidence level). 
The $\sim 9 \times 10^{17}$~atoms of \K\ in the source have an activity of $\sim 16$~Bq, equivalent to two bananas~\cite{hoeling1999going}, and the source is 5.1(9)~$\mu$m thin to allow the X-rays to escape from it. The source rests directly in front of the SDD and is centered inside MTAS.}

MTAS is a 1-tonne array of NaI scintillators with \amend{$\sim 98\%$} tagging efficiency for the 1.46~MeV $\gamma$-rays of interest (Table~7 in~\cite{stukel2021novel}).
Data with the \K\ source were acquired over 44~days.  
Following offline determination of coincidences between the SDD and MTAS  for three nominal time windows of $(1, 2, 4)\ \mu \text{s}$, SDD pulses were fit and energy calibrations were performed.  To avoid biases during the analysis, the anti-coincident SDD spectrum had been  blinded from (0.88--1.4)~keV (silicon escape peak region) and (2.0--3.8)~keV (electron capture signal region) while cuts and analysis methods were established.

Over the course of the run, two types of SDD instabilities appeared: gain drops due to voltage supply failures that were reset by an operator, and noise bursts attributed to power fluctuations in the lab.  Both can be identified in the SDD energy range below the blinded region; after this cut, $ 76\%$ of the livetime remains.  Minor gain drifts in the SDD were corrected by tracking the coincidence X-ray lines.

Gain drifts of a few percent were also observed in MTAS. A Kolmogorov-Smirnov test comparing the arrival times of the open anti-coincident and coincident SDD events in the 2--5~keV region returned a p-value of 0.63.  This implies both data sets are consistent with the same underlying time distribution, and is consistent with the tagging efficiency of MTAS not changing over the run.  In addition, before opening the data set,  five time regions with MTAS gain changes were identified.  After opening the data, the full $\rho$ analysis (detailed in later sections) was carried out on each of these regions, and the values were compared.  A fit to a single common value of $\rho$ yielded a $\chi^2$ of  3.6 for 4 degrees of freedom, i.e., $p = 0.45$, implying the data are consistent with a constant tagging efficiency.  Consistent results were observed over all three coincidence windows.

\subsection{Physical phenomena visible in the SDD and MTAS  spectra}

\begin{figure}[ht]
  \centering
  \includegraphics[width=1.0\linewidth]{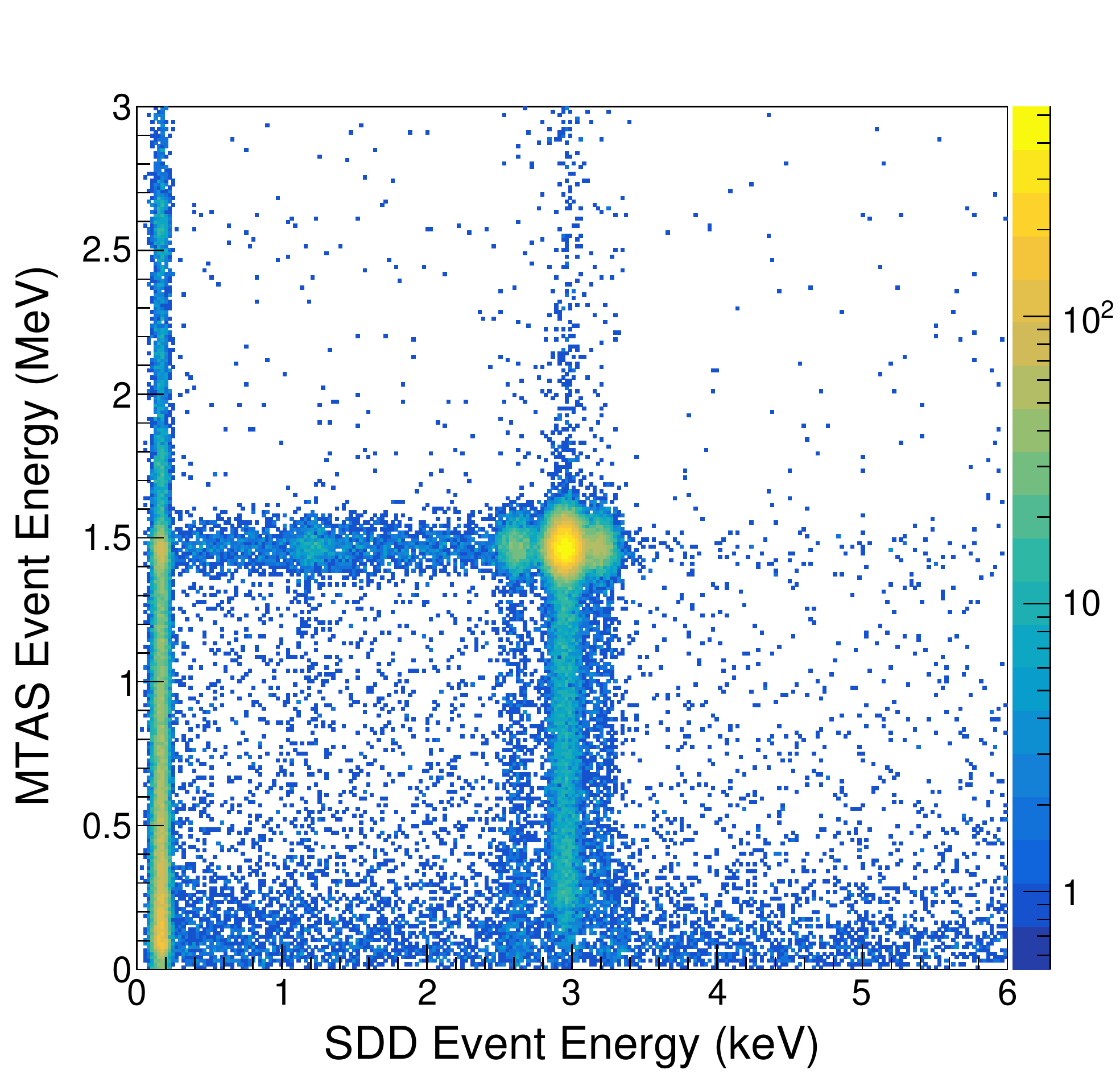}
	\caption{\label{Fig:K_40_1us_Coincidence_Spectrum.pdf}\K\ coincidence spectrum between the SDD and MTAS. SDD and MTAS data obtained with the enriched \K\ source using a nominal 2~$\mu$s coincident window is binned into a 250 $\times$ 250 grid over the displayed energy ranges. }
\end{figure}

A coincidence spectrum of the \K\ source with a 2~$\mu$s nominal window can be seen in Fig.~\ref{Fig:K_40_1us_Coincidence_Spectrum.pdf}, which bins events by SDD and MTAS energies, resulting in various bands and peaks. The foci are near (\keV{3}, \MeV{1.46}), corresponding to \ECStar\ decays involving X-ray detection in the SDD and complete capture of the de-excitation $\gamma$-ray in MTAS. Additional features in  Fig.~\ref{Fig:K_40_1us_Coincidence_Spectrum.pdf} involve partial energy depositions of \ECStar\ decay products. The projection of this figure, with nonzero MTAS energy, onto the SDD energy, is the coincident SDD data spectrum in Fig.~\ref{Fig:K40_Anti_vs_Coinc_Histogram.pdf}.

In addition to the Ar signal, characteristic X-rays of Cl, K, and Ca  are seen in the signal region (see Table~\ref{tab:fluo_counts} for the number of events, as well as Fig.~\hl{2} of~\cite{prl}). 
K and Cl atoms are numerous enough in the source to be fluoresced by source $\beta^-$. Most $\beta^-$-fluoresced events are anti-coincident, with a small coincident contribution primarily from the $\beta^-$ which make it into MTAS. 
Relative coincident and anti-coincident K counts are consistent with this description. Though Cl is also fluoresced by $\beta^-$, Ar X-rays (generally from \ECStar) are energetic enough to fluoresce Cl, contributing additional coincident events.

\begin{table}[ht]
    \centering
    \caption{\label{tab:fluo_counts}Fluorescence counts observed in SDD spectra. Fluorescence may occur via various sources, described in the text, which affect the coincident sorting of the resulting X-ray detection. The quoted values are obtained using a \mus{2} window between the SDD and MTAS.}
    \begin{ruledtabular}
    \begin{tabular}{l D{,}{\times}{-1} D{,}{\times}{-1}}
        Element    &   \multicolumn{1}{c}{Coincident counts}  &   \multicolumn{1}{c}{Anti-coincident counts}    \\ 
        \hline
        Cl    &    3.48(2) , 10^3    &   8.85(5) , 10^3    \\
        K  &   2.9(3), 10^2 &   5.89(7) , 10^3   \\
        Ca &  1.0(5) , 10  &   1.36(5) , 10^3
    \end{tabular}
    \end{ruledtabular}
\end{table}

The Ar and Ca atoms in the source  come from the slow  decay of \K.  There are therefore 9--10 orders of magnitude fewer Ar and Ca atoms than K or Cl ones in the source. Based on the limited K fluorescence visible in KDK data, fluorescence of an Ar or Ca atom  caused by something other than the atom itself  is highly negligible. 

Interestingly, there is a small contribution of characteristic Ca X-rays visible in the anti-coincident spectrum. Having ruled out external fluorescence, a product of the \K\ $\rightarrow $ \Ca\ decay itself must produce these events; the decay to Ca must occasionally involve \emph{self-fluorescence} of the daughter via the produced $\beta^-$. The exact low-order processes involved in such interactions are beyond the scope of this work, and we note that the Ca contribution has no effect on analysis of the Ar X-rays.

Outside of the SDD signal region, an additional presentation of source \ECStar\ decays is contained at (\keV{1.2}, \MeV{1.46}). This silicon escape peak is formed in the event that a source X-ray fluoresces a detector Si atom prior to detection. The remaining energy of the source X-ray is equivalent to the difference between its initial energy and that of the fluoresced Si X-ray (\keV{1.7}), resulting in a detectable \keV{1.2} event.

Lastly, though it is possible for source $\gamma$-rays to deposit energy in the SDD, this is modelled to occur in $<0.5\%$ of cases. This contribution is considered in the systematic analysis described further, though the effect is negligible relative to the statistical error of our measurement. $\beta^+$ from the source would eventually provide a small contribution to the continuous SDD background, though over the KDK runtime any such events are negligible due to the minute branching ratio of this mode.

\subsection{Coincident and anti-coincident events}\label{sec:CoinUnco}

In the KDK dataset, we expect a total of $\sigma^*$ counts from \ECStar\ decays, and $\sigma$ counts from \EC\ decays present in our signal (Ar X-ray) peaks. These are expanded to
\begin{align*}
    \sigma^*  & = AT\ \text{\PECStar }  \ P_K^* \omega_K \  \eta \ ( 1- \eta_\gamma ) \nonumber
    \\
    \sigma  & = AT \ \text{\PEC } \  P_K \omega_K \  \eta ,
\end{align*}
where $AT$ are total source decays over the run duration. K-shell capture probabilities $P_K^* = 0.7564(4), P_K = 0.8908(7)$ using Betashape code V.2.2~\cite{mougeot2017betashape} differ for the two modes, though the fluorescence probability $\omega_K$ is the same. Both modes emit the same X-rays, whose detection probability in the SDD is $\eta$. The $\gamma$-ray accompanying the \ECStar\ decay could deposit some energy in the SDD, with a small probability, thus shifting the event out of the signal region. From simulations, we estimate this probability to be $\eta_\gamma = 0.0048(48)$.

Various factors will sort $\sigma$ and $\sigma^*$ into coincident and anti-coincident events; the main ones are:
\begin{enumerate}
    \item{The efficiency with which the $\gamma$s from \ECStar\ are tagged by MTAS is not perfect.  We have previously studied this parameter $\amend{\epsilon}$ and determined it with high precision to be $\amend{\epsilon}=0.9792 \pm 0.006$ at a 2~$\mu$s coincidence window~\cite{stukel2021novel} using data and Geant4~\cite{collaboration2003geant4} simulations.  This will reduce the expected number of \ECStar\ events that are properly tagged to $ \sigma^* \epsilon $.}
    
    \item{Conversely, the non-perfect tagging efficiency will lead to $\sigma^* (1 - \epsilon )$ EC$^{*}$s being untagged (false positives).}
    
    \item{In addition, some \EC\ events may be in spurious coincidence with the MTAS background; the probability that this occurs in the $\mathcal{O}(\mu \text{s})$ coincidence window is $\beta_M \bar{t} << 1$~\cite{stukel2021novel} (false negatives).}
    
    \item{The remaining \EC\ events are anti-coincident.}
\end{enumerate}
These subsets of signal counts are summarized below:
\begin{align*}
    \Sigma^*  & = 
        \overbrace{
            \sigma^* \epsilon 
        }^\text{1. Observed \ECStar}
        + 
        \overbrace{
            \sigma \beta_M \bar{t} 
        }^\text{3. False negatives} \nonumber
    \\
    \Sigma & =  
        \underbrace{
            \sigma^* (1 - \epsilon ) 
        }_\text{2. False positives}
        + 
        \underbrace{
            \sigma ,
        }_\text{4. Observed \EC}
\end{align*}
where $\Sigma^*$ are expected coincident counts, and $\Sigma $ are expected anti-coincident counts. We introduce a parameter for total signal counts, $\nu \equiv \sigma^* + \sigma $, and a useful term:
\begin{equation*}
    \rho^\prime \equiv \frac{\sigma }{\sigma^* } = \rho \frac{P_K}{P_K^*} \frac{1}{(1 - \eta_\gamma )}.
\end{equation*}

Using these two new expressions, we obtain the final likelihood terms relating $\rho $ to expected coincident and anti-coincident counts:
\begin{align}
    \Sigma^* & =\frac{\nu }{ 1 + \rho^\prime } \left( \epsilon + \rho^\prime \beta_M \bar{t} \right) \nonumber
    \\
    \Sigma  & = \frac{\nu }{ 1 + \rho^\prime } \left( 1 - \epsilon + \rho^\prime  \right) .
    \label{eq:exp_coinc_uncoinc_counts}
\end{align}

\subsection{Likelihood method}

SDD data are sorted by coincidence, and both subsets are fit simultaneously through minimization of the sum of the  negative-log likelihoods:
\begin{equation*}
    -\ln \mathcal{L}   = - (\ln \mathcal{L}_{coin} + \ln \mathcal{L}_{anti})
\end{equation*}
Each of the two terms is a binned Poisson  likelihood ratio~\cite{BAKER1984437}:
\begin{equation*}
     - \ln \mathcal{L}_{j}  =     
    \sum_i \left[ f(x_{i, j};\boldsymbol{\theta})  - n_{i, j} + n_{i, j} \ln \biggl( \frac{n_{i, j}}{   f(x_{i, j};\boldsymbol{\theta} ) } \biggr) \right].
\end{equation*}
Above, index $j$ represents either coincident or anti-coincident data, $n_{i, j}$ are total events in bin $i$ and $f(x_{i, j})$ are the model-predicted counts in that bin.
In addition to providing estimators of the parameters and confidence intervals, this technique returns goodness-of-fit~\cite{BAKER1984437}.  Some of the parameters in $\boldsymbol{\theta}$, like our main one $\rho = {\PEC} / {\PECStar}$, and like those pertaining to the shapes of the lines,  are shared between the coincident and anti-coincident data, while others are not.

The spectra contain several  fluorescent contributions (Cl, K, Ca).  For each, we model the associated \Kalpha\ and \Kbeta\ X-rays with Gaussian distributions, the means of which are fixed to known values~\cite{be_table_2010}. For each such \Kalpha\ + \Kbeta\ pair, the free Gaussian width is shared across the two peaks, and is the same in both coincident and anti-coincident spectra. A parameter for total \Kalpha\ + \Kbeta\ counts, not shared across the coincidence-sorted data, is left free. The ratio of intensities $I_{K_\alpha}/ I_{K_\beta}$ is generally fixed to values in~\cite{SCHONFELD1996527}.

The continuous background model, consisting of decaying exponential and flat components, has all associated parameters left free. These parameters are not shared across the coincident and anti-coincident data, as this background contribution has a different shape in each subset.

The Ar \Kalpha\ and \Kbeta\ X-rays of interest are modelled in a similar manner as the fluorescence lines. In order for these components to directly inform $\rho$, we insert the expression for expected coincident and anti-coincident Ar counts of Eq.~\eqref{eq:exp_coinc_uncoinc_counts} directly into the likelihood. This introduces free parameters $\rho $ and total Ar counts, along with fixed terms including efficiencies, as described earlier in Sec.~\ref{sec:CoinUnco}.

We note that the result $\rho $ is stable to the choice of fixing or freeing the ratio of \Kalpha\ to \Kbeta\ intensities in the Ar and fluorescent components. Modelling of the latter generally has a negligible effect on the result, which is informed only by the Ar lines. An additional test modelled the signal X-rays with a Voigt profile~\cite{wertheim1974determination}, which yields essentially identical results since it approaches the limiting Gaussian case.

\begin{figure}[ht]
  \centering
    \includegraphics[width=\linewidth]{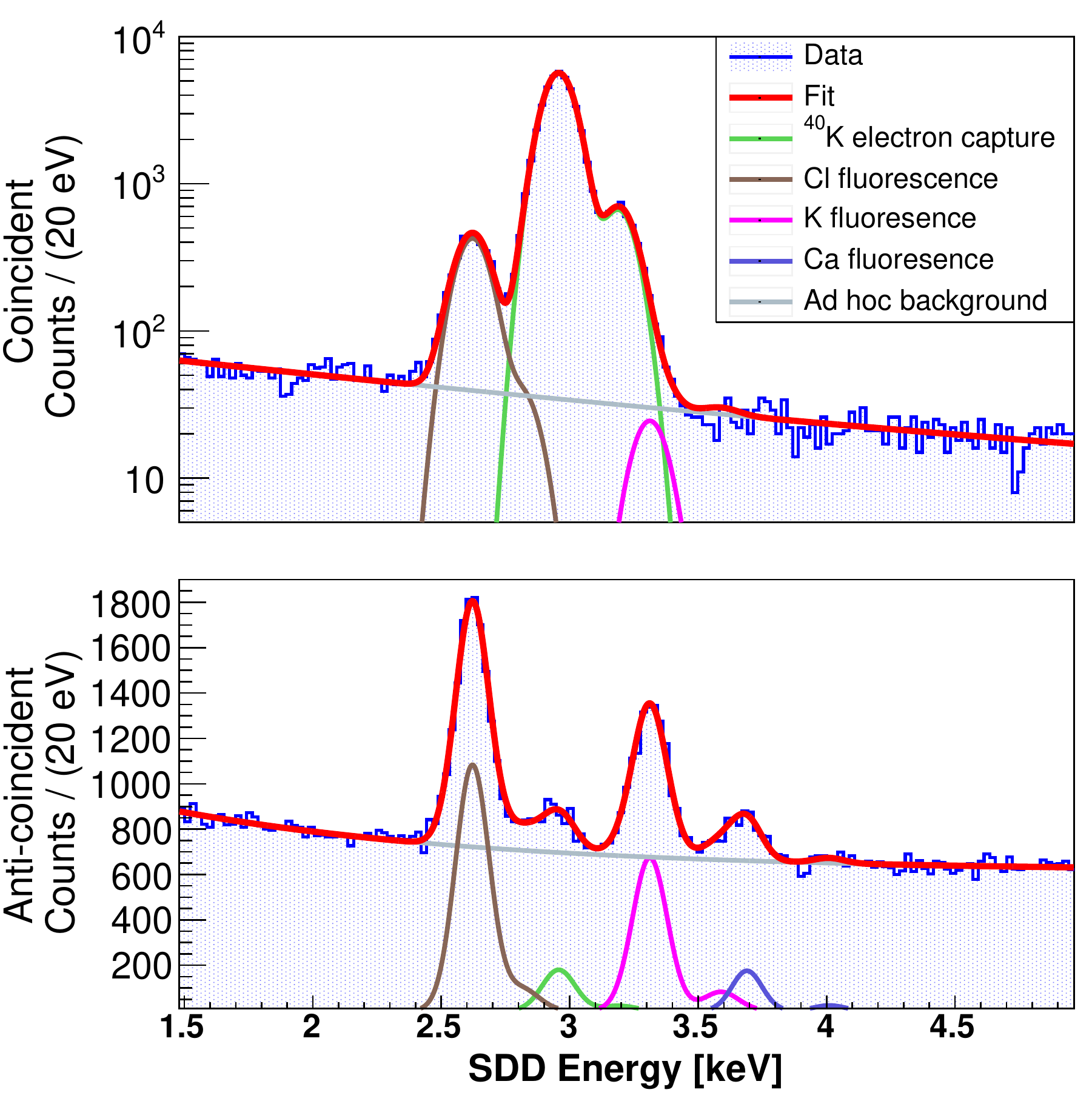}
	\caption{\label{Fig:K40_Anti_vs_Coinc_Histogram.pdf}SDD coincidence and anti-coincidence spectra. Results of simultaneous fit to coincident (top) and anti-coincident (bottom) SDD spectra at a \mus{2} coincidence window. Signal counts are shown in green. Various fluorescence peaks and an exponential background model are included. The total minimization has an associated goodness-of-fit of $p=0.4$. }
\end{figure}

Initial opening of the data led to a value for $\rho$ of $0.008 \stackrel{stat}{\pm} 0.002$ on the single chosen  energy range before evaluation of systematic errors.
A thorough analysis of systematics  leads to our reported value of $\rho =\rhoRes$.
An example of a fit is shown in Fig.~\ref{Fig:K40_Anti_vs_Coinc_Histogram.pdf}.

\subsection{Systematics}

The systematic errors pertinent to the experiment fall into two categories: fit characteristics, and physical limitations. The choice of fit range (canonical fit range is 1.5--5.0~keV) pertaining to the mathematical background model is the dominant source of systematic error. The dominant physical error arises from the imperfect $\gamma$-ray tagging efficiency of MTAS. Table~\ref{tab:syst_errors} contains a summary of each source of error and its effect.

\begin{table}[ht]
    \centering
    \caption{\label{tab:syst_errors}Systematic, \amend{68\% C.L.} errors on $\rho$. All sources of error, described in the text, are smaller than the statistical (\ratiostat). }
    \begin{ruledtabular}
    \begin{tabular}{l D{,}{\times}{-1}}
        Source  &  \multicolumn{1}{c}{Systematic Error}    \\
        \hline 
        Fit range     & \amend{9} , \amend{10^{-4} }   \\
        MTAS $\gamma$-ray-tagging efficiency     &  5 , 10^{-4}  \\ 
        Binning &   1 , 10^{-4}   \\
        SDD $\gamma$-ray-tagging efficiency    &   4 , 10^{-5} \\
        K-shell capture probabilities    & 8 , 10^{-6}   \\
    Expected MTAS background counts &  3 , 10^{-6}     \\
        \end{tabular}
    \end{ruledtabular}
\end{table}

\begin{figure}[ht]
    \centering
    \includegraphics[width=\linewidth]{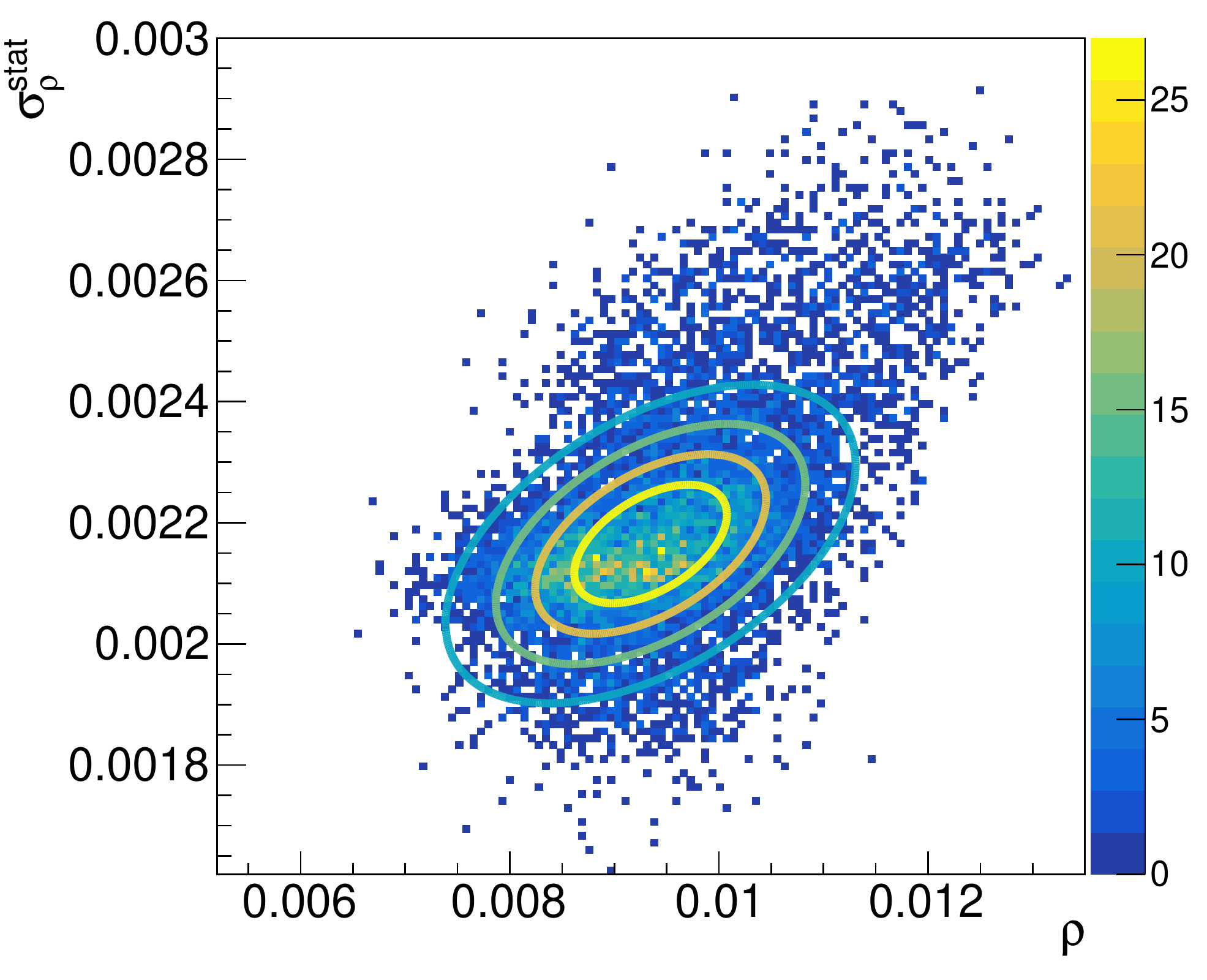}
    \caption{ Distributions of $\rho $ and its statistical error from systematic checks.
    The value of $\rho$ and its associated statistical error $\sigma_\rho^\text{stat}$ obtained from 10,000 systematic-varying fits are displayed. The width of the $\rho $ distribution is equivalent to the systematic error. Contour lines of the fit function correspond to the colour bar on the right.}
    \label{fig:syst_check_rho_rho_error}
\end{figure}

To account for possible covariances between parameters which contribute to the overall systematic error, their effect is tested simultaneously by randomly drawing their value prior to a likelihood fit. Fit characteristics are drawn from a uniform distribution over the considered range. Physical parameters are drawn from a Gaussian whose width corresponds to the parameter's known uncertainty. This process is repeated 10,000 times to obtain a distribution of $\rho $, whose mean corresponds to our final measurement, and width is the systematic error, as shown in Fig.~\ref{fig:syst_check_rho_rho_error}.

Though the systematic check described above uses a background model consisting of exponential and flat components, other models have been explored in depth. We have tested polynomials up to the third degree over various ranges, and find them to either be too simple or to find unphysical minima and maxima in the data. Overall, we find other models to either be too simple, find unphysical artefacts, or supply extraneous degrees of freedom to the canonical (exponential) case.

Lastly, our analysis was performed at 3 coincidence windows: $(1, 2, $ and $4)~\mu \text{s}$, for which we find consistent results of $\rho = (0.0091, 0.0095, 0.0095)$ respectively, all with the same error $\stackrel{stat}{\pm} 0.0022 \stackrel{sys}{\pm} 0.0010$.
These values are consistent with the value initially found upon opening the data and reported in the previous section.
Ar counts obtained in the systematic check at \mus{2} are summarized in Table~\ref{tab:Ar_counts_summary}.

\begin{table}[ht]
    \centering
    \caption{\label{tab:Ar_counts_summary}\K\ $\rightarrow$ \Ar\ electron capture decays visible in \mus{2} KDK data. The total $\sim 3$~keV signal counts pertaining to this transition are sorted by coincidence, with ground-state decays contributing primarily to the anti-coincident subset.  Notations in parentheses refer to Sec.~\ref{sec:CoinUnco}.}
    \begin{ruledtabular}
    \begin{tabular}{l D{,}{\times}{-1}}
        Component & \multicolumn{1}{c}{Visible counts} \\
         \hline
        Total Ar ($\Sigma^* + \Sigma$)   & 4.78 , 10^4   \\
        Coincident Ar ($\Sigma^*$) &   4.63 , 10^4  \\
        Anti-coincident Ar   ($\Sigma$) &   1.50 , 10^3  \\
        \EC\ decay ($\sigma$)  &   5.00 , 10^2
    \end{tabular}
    \end{ruledtabular}
\end{table}

\section{Implications}

The following re-evaluation of the \K\ decay scheme using our measurement affects a variety of fields, including nuclear structure theory and geochronology. This is discussed below, illustrating impacts on neutrinoless double-beta decay of ${}^{48}$Ca and the thermal history of the Acapulco meteorite.

\subsection{Construction of the decay scheme of \texorpdfstring{\K}{40K}}

Constructing the full decay scheme for \K\ requires 4 parameters.  Two of these are partial decay constants for the $\beta^-$ and \ECStar\ branches, $\lambda^- = 0.4904 \pm 0.0019$~Ga$^{-1}$ and $\lambda^* = 0.05646 \pm 0.00016 $~Ga$^{-1}$ (where Ga denotes $10^9$~yr), taken from the most recent data evaluation  (Sec.~5.2 of~\cite{kossert2022activity}).  These absolute measurements are independent of \EC, though they  depend on factors like the precise \K\ content of the source, and the efficiency of the detectors. Generally speaking, the other two parameters used are an experimental determination of \PBetaPlus /\PBetaMin ~\cite{engelkemeir_positron_1962} and a theoretical value for \PEC /\PBetaPlus~\cite{mougeot_improved_2018}.  The former directly leads to $\lambda^+ = \frac{\PBetaPlus}{\PBetaMin} \lambda^-$, while the latter then provides $\lambda^0 = \frac{\PEC}{\PBetaPlus} \lambda^+$.  From this complete set of partial decay constants, the total decay constant can be obtained: $\lambda = \sum_i \amend{\lambda_i}$.  Partial and total halflives are then determined  ($T_i = \frac{\ln 2}{ \lambda_i}$). Lastly, branching ratios ($P_i = \frac{\lambda_i}{\sum_j \lambda_j}$) are obtained.

Our novel measurement provides an additional, experimental, value: \PEC/\PECStar.   In conjunction with $\lambda^*$, it leads directly to $\lambda^0 = \frac{\PEC}{\PECStar} \lambda^*$. 
The value of $\lambda^0$ is the same, within uncertainties, for various commonly-used sets of decay constants~\cite{kossert2022activity,min2000test,be_table_2010}.
We keep $\lambda^-$, as before.  To complete the set of 4 parameters with $\lambda^+$, we need to choose  between \PBetaPlus /\PBetaMin\  and \PEC/\PBetaPlus.  This choice only affects the decay scheme at the level of the small $\beta^+$ branch which varies by a factor of 2, as Table~\ref{tab:New_Decay_Scheme_Measurements} shows.  This discrepancy advocates for further experimental and theoretical work on the $\beta^+$ branch.

\begin{table}[ht]
    \centering
    \caption{\label{tab:New_Decay_Scheme_Measurements}Re-evaluation of the \K\ decay scheme. Branching ratios $I$ and total half life $T_{1/2}$ are calculated from our measurement of $\rho=\text{\PEC / \PECStar}$, evaluation of measured partial \ECStar\ and $\beta^-$ half lives~\cite{kossert2022activity}, and either the measured relative $\beta^+/\beta^-$ feeding~\cite{engelkemeir_positron_1962} or the predicted value of \PEC/\PBetaPlus~\cite{mougeot_improved_2018}.  The choice only affects the $\beta^+$ branching.}
    \begin{ruledtabular}
    \begin{tabular}{l D{.}{.}{-1} l}
        Quantity & \multicolumn{1}{c}{Value}  &  Uncertainty \amend{(68\% C.L.)}\\
        \hline
        \PEC\ (\%) & \IECBranchingPer  & $\pm  \IECBranchingStatPer(\text{stat}) \pm \IECBranchingSystPer(\text{syst})  $ \\
        \PECStar\ (\%) & \IECStarBranchingPer  & $\pm \IECStarBranchingErrPer $  \\
        \PBetaMin\ (\%) & \IBetaMinBranchingPer  & $\pm \IBetaMinBranchingErrPer $  \\
        \PBetaPlus\ (\%) (expt) & \IBetaPlusExptBranchingPer  & $\pm \IBetaPlusExptBranchingErrPer $  \\
        \PBetaPlus\ (\%) (theory) & \IBetaPlusTheoryBranchingPer  & $\pm \IBetaPlusTheoryBranchingErrPer $  \\
        $T_{1/2}$ (Ga) &   \ThalfTotGa  &   $\pm \ThalfTotErrGa $
    \end{tabular}
    \end{ruledtabular}
\end{table}

\subsection{\label{sec:Theory}Nuclear shell-model calculations}

 We obtain a theoretical estimate for the third-forbidden unique \PEC\ transition, that is complementary to the main experimental result, as described further. Moreover, the (in)frequency of this decay informs the extent to which such forbidden modes are suppressed, which is applicable to calculations of neutrinoless double-beta decay half-lives including that of $^{48}$Ca. This commonly overlooked suppression, quantified as quenching of the weak axial-vector coupling, can significantly increase calculated $0\nu\beta\beta $ half-lives.
 
\begin{figure}[ht]
  \centering
  \includegraphics[width=1.0\linewidth]{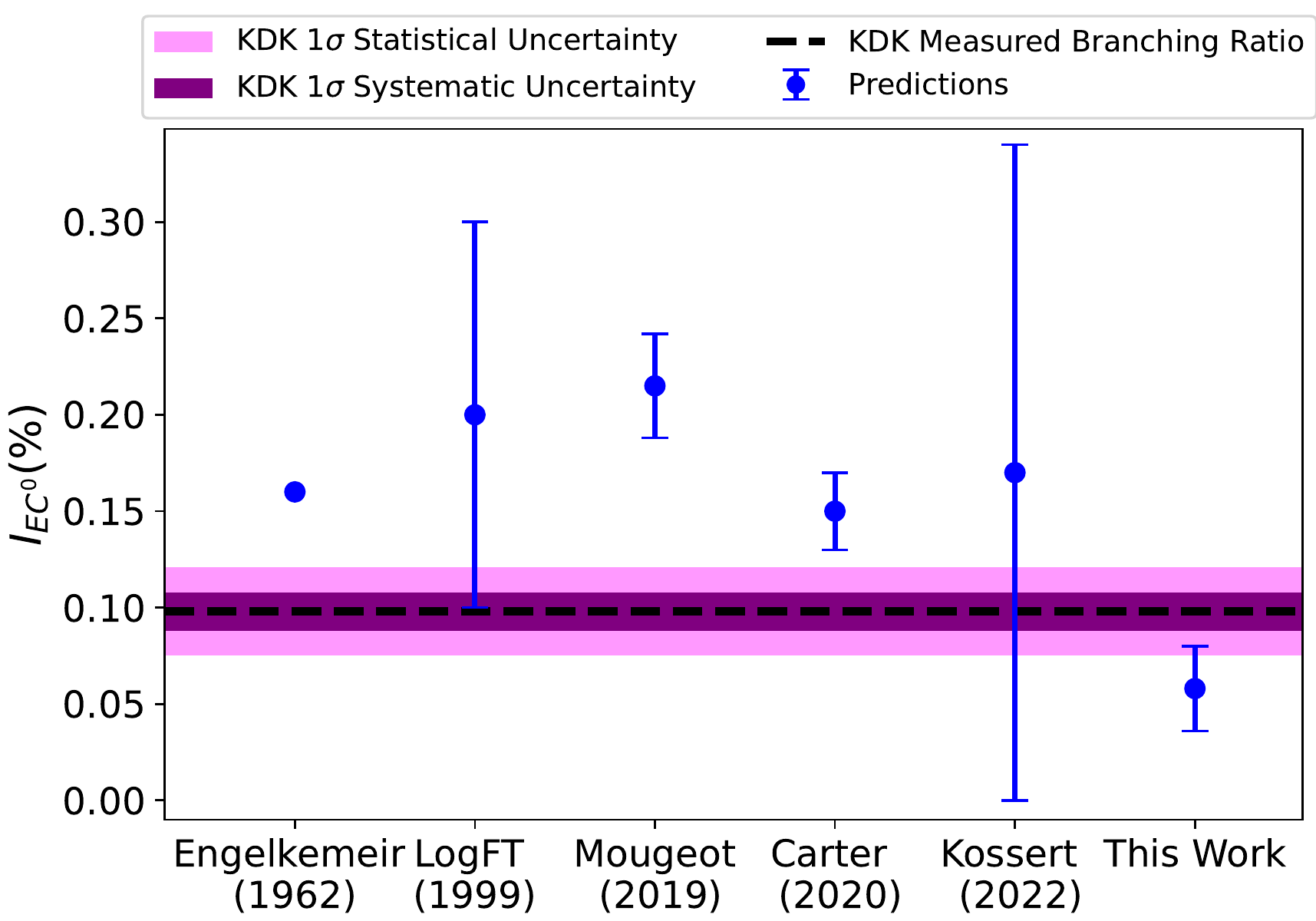}
	
    \caption{\label{Fig:BREC_History_W_KDK_Sensitivity.png}Measured ground-state electron-capture branching ratio compared to predictions from Engelkemeir (1962) \cite{engelkemeir_positron_1962}, $\log ft$ (1999)~\cite{gove_log-f_1971,noauthor_logft_nodate}, Mougeot (2019)~\cite{mougeot_improved_2018}, Carter (2020)~\cite{carter2020production}, Kossert (2022)~\cite{kossert2022activity}~and this work (Sec.~\ref{sec:Theory}). \amend{All error bars correspond to a 68\% C.L., with no uncertainty reported for Engelkemeir}.}
\end{figure}

 We calculate a value of $\PEC = 0.058 \pm 0.022$  using the Behrens-B\"uhring formalism (\cite{behrens1982} for the full theory and~\cite{haaranen2017} for a streamlined version) with the nuclear matrix elements calculated in the shell-model framework using the code NuShellX@MSU~\cite{nushellx} with the Hamiltonian \emph{sdpfk}~\cite{sdpfk}. It is consistent within uncertainties with a value obtained prior to the experiment. Our theoretical estimate is compared to other predictions along with the KDK measurement of this work in Fig.~\ref{Fig:BREC_History_W_KDK_Sensitivity.png}.

Using this Hamiltonian the half-lives of the three decay branches, $\beta^-$, \ECStar, and \EC, could be calculated. The corresponding decay amplitudes are proportional to the weak axial-vector coupling $g_{\rm A}$, the value of which is known to be quenched for a wide range of nuclear masses~\cite{Suhonen2017}. Here the effective value of it can be determined by comparison of the computed and experimental half-lives, giving for the first-forbidden unique transition \ECStar\ the value $g_{\rm A}^{\rm eff}=0.34$ and for the two third-forbidden unique transitions a value $g_{\rm A}^{\rm eff}=0.43$ for $\beta^-$ branch and $g_{\rm A}^{\rm eff}=0.53$ for the \EC\ branch. These values of $g_{\rm A}^{\rm eff}$ are very well in line with the values $g_{\rm A}^{\rm eff}=0.43$ -- $0.66$ obtained in the mass range $A=74$ -- $126$ for the $2^{-}\leftrightarrow 0^{+}$ first-forbidden unique $\beta$ and EC transitions in the framework of the proton-neutron quasiparticle random-phase approximation~\cite{EJIRI201427}. These results suggest that the forbidden contributions to the nuclear matrix elements of the \znbb\ decay are very much suppressed and the resulting  half-lives are much longer than expected based on the bare value $g_{\rm A}^{\rm bare}=1.27$ of the axial-vector coupling. 
Using values in~\cite{Kortelainen_2004}, we expect the quenching of this axial-vector coupling strength to increase the neutrinoless double-beta decay half-life of $^{48}$Ca by a factor of $7^{+3}_{-2}$.

\subsection{Effect on geochronology}

The branched decay of \K\ to \Ar\ is the basis of K/Ar dating and its variant the \ArAr\ technique. The ubiquity of K and the $\sim$1.27~Ga half life of \K\ make these geochronometers some of the most useful and versatile isotopic methods available to date geological samples. 

The age equation for the \K/\Ar\ isotope system is:
\begin{equation*}\label{Eqn:K_Ar_Age}
    t_{\small\KAr} = \frac{1}{\lambda_T} \ln\left[\frac{N_{^{40}\text{Ar}}}{N_{^{40}\text{K}}}\frac{\lambda_T}{\lambda_\text{Ar}} + 1\right],
\end{equation*}
where $t_{\small\KAr}$ is the age of the sample, $N_x$ is the number of atoms of a given isotope in the sample, $\lambda_T$ is the total decay constant of \K\ and $\lambda_\text{Ar}$ is the decay constant of \K\ to \Ar, including \ECStar, \EC\ and $\beta^+$ branches.
In the K/Ar technique, the ratio of \K\ to \Ar\ isotopes is measured, leading to an estimation of the age of the sample.  Precision of this technique currently reaches 0.5\%~\cite{mcdougall2011calibration}.

In the \ArAr\ technique, neutron activation is used to transmute $^{39}$K to $^{39}$Ar.  This allows mass-spectrometric measurements of the parent proxy ($^{39}$Ar) and the daughter (\Ar) on the same sample. The efficiency of the activation is hard to estimate, therefore the activation is generally carried out in parallel on a reference sample of known age, providing the age of the sample relative to the reference. The age of the reference must be established independently of the \ArAr\ technique. Commonly used references include Fish Canyon ($t_m = 28.201 \pm 0.023$~Ma for a 68\% CL)~\cite{kuiper2008synchronizing}. The age of the sample is then:
\begin{equation*}\label{Eqn:Ar_Ar_Age}
    t_{\small\ArAr} = \frac{1}{\lambda_T} \ln\left[\frac{N_{^{40}\text{Ar}}}{N_{^{39}\text{Ar}}}J+1 \right],
\end{equation*}
where the irradiation parameter ($J$) is  given by: 
\begin{equation*}\label{Eqn:Monitor_Flux}
    J = \frac{e^{\lambda_T t_m}-1}{M_{^{40}\text{Ar}}/M_{^{39}\text{Ar}}},
\end{equation*}
where $M_{^{40}\text{Ar}}/M_{^{39}\text{Ar}}$ is the measured isotopic ratio in the monitor.  Unlike K/Ar, the \ArAr\ technique no longer depends on the argon branching ratio. This technique currently reaches precisions of 0.1\%~\cite{niespolo2017intercalibration}.

Certain sets of decay constants widely used in the geological community question the existence of, or ignore, the \EC\ decay branch~\cite{min2000test,renne2010joint}. Using data in Table~\ref{tab:Geo_Decay_Constants_main}, we illustrate how adding the \EC\ branch to various sets of decay constants~\cite{min2000test,kossert2022activity} affects K/Ar dates throughout geologic time in Fig.~\hl{4a} of~\cite{prl}.  
In Fig.~\hl{4b} of~\cite{prl} we display the impact of this reevaluation on the commonly-used Fish Canyon sanidine standard, when the K/Ar age is recalculated from the  \K /\Ar\ ratio~\cite{jourdan2007age} and various decay constants~\cite{min2000test,kossert2022activity,steiger1977subcommission}.
Using these updated Fish Canyon ages with the same set of decay constants, we recalculate the \ArAr\ ages of the Acapulco meteorite~\cite{renne200040ar} in Fig.~\hl{4c} of~\cite{prl}.
Also shown is the Pb/Pb age for phosphates from Acapulco~\cite{gopel2010thermal}, updated to include uncertainties in the uranium isotopic composition~\cite{GOLDMANN2015145}, and the uranium decay constants~\cite{jaffey1971precision} ($4.555 \pm 0.005$~Ga). 

\begin{table*}[ht]
    \centering

    \caption{\label{tab:Geo_Decay_Constants_main}Effect of adding $\EC$ to a commonly used set of decay constants~\cite{min2000test} and the latest evaluation~\amend{(Sec.~5.2 of~\cite{kossert2022activity})}. $\lambda_T$ is the total decay constant of \K , and $\lambda_\text{Ar}$ is its partial decay constant to \Ar .}
    \begin{ruledtabular}
    \begin{tabular}{l D{,}{\pm}{-1} D{,}{\pm}{-1}}
      & \lambda_T , 1\sigma \text{ (Ga}^{-1}\text{)} & \lambda_\text{Ar} , 1\sigma \text{ (Ga}^{-1}\text{)} \\\hline
    Min et al.~\cite{min2000test} & 0.546 , 0.005 & 0.0580 , 0.0007     \\
    KDK with Min et al.~\cite{min2000test}  &   0.547 , 0.005 &  0.0585 , 0.0007 \\
    Kossert et al.~\cite{kossert2022activity} & 0.5468 , 0.0019 & 0.0564 , 0.00016     \\
    KDK with Kossert et al.~\cite{kossert2022activity}  &   0.5474 , 0.0019 &  0.0570 , 0.00021 \\
    \end{tabular}
    \end{ruledtabular}
\end{table*}

\begin{figure}[ht]
 \centering

             \includegraphics[width = 1\linewidth]{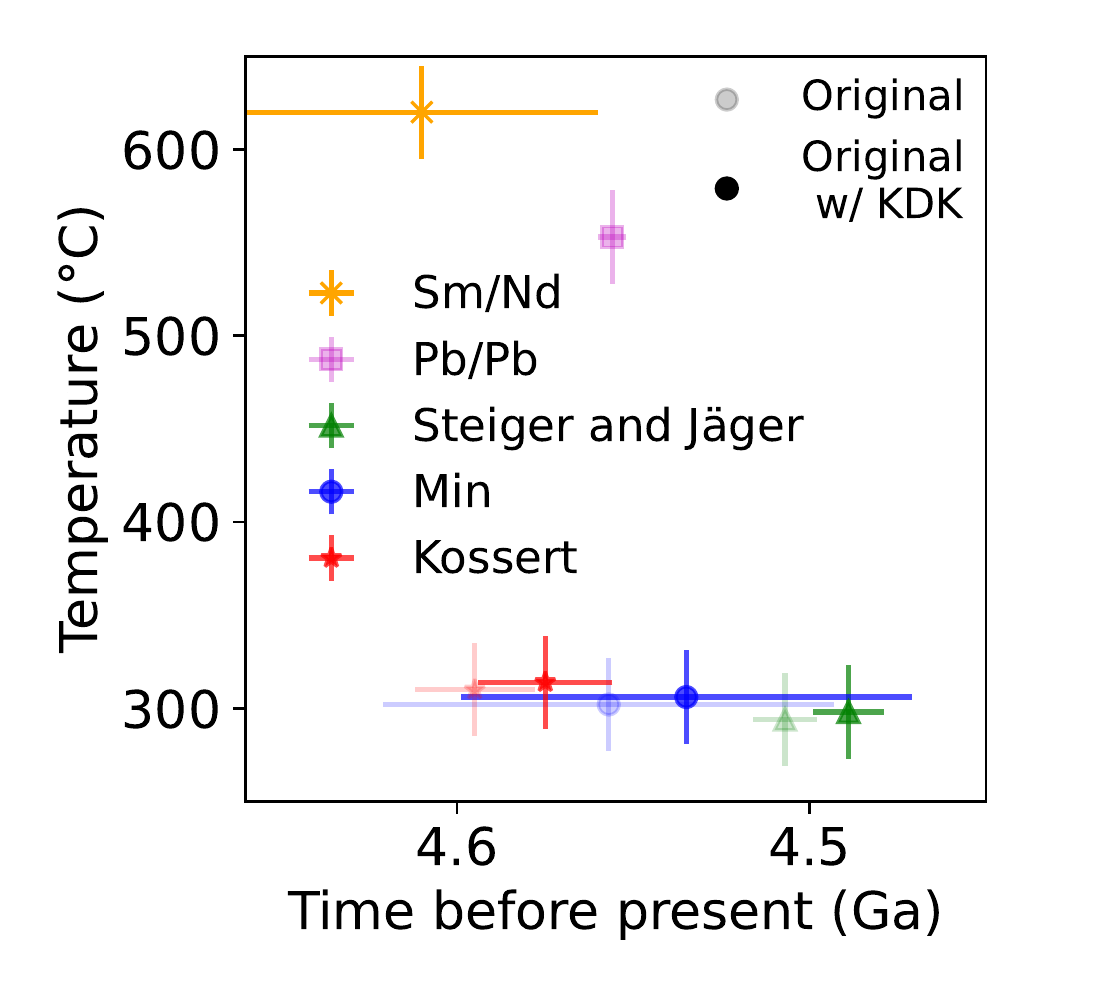}
 
    \caption{Effect of KDK on the thermal history of the Acapulco meteorite. \ArAr\ ages of the Acapulco meteorite~\cite{renne200040ar} calculated for various sets of decay constants~\cite{steiger1977subcommission,min2000test,kossert2022activity} (Fig.~\hl{4c} in \cite{prl}) are plotted along with the corresponding estimated closure temperature of $300 \pm 25$~$^\circ$C (temperatures shifted for visibility). Age-temperature data obtained using Pb/Pb (updated from~\cite{gopel2010thermal}) and Sm/Nd~\cite{prinzhofer1992samarium} also shown. \amend{For reference, the Pb/Pb age of the oldest dated solid in the solar system, a calcium-aluminum inclusion from the CV chondrite Efremovka, is $4567.35 \pm 4.65$~Ma~\cite{Connelly2012}.} \amend{All uncertainties correspond to a 68\% confidence level, and include decay constant uncertainties.}
    \label{fig:ThermalHistory_main}}
\end{figure}

Fig.~\ref{fig:ThermalHistory_main} shows the thermal history of the Acapulco meteorite using the various \ArAr\ ages, along with those from Pb/Pb and Sm/Nd~\cite{prinzhofer1992samarium}, with closure temperatures, and updated uncertainties for Sm/Nd, from~\cite{renne200040ar}.
Pb/Pb and Sm/Nd ages are statistically indistinguishable from \ArAr\ ages calculated using recent decay constants~\cite{min2000test,kossert2022activity} updated by our measurement of \EC\ ($\chi^2$ fits to a common age respectively yield $p=0.37$ for updated Kossert~\cite{kossert2022activity} and $p=0.51$ for updated Min~\cite{min2000test}), consistent with rapid cooling.
The KDK measurement itself tends to decrease \ArAr\ ages and therefore reduce apparent cooling rates.  The systematic nature of this change may affect studies of past heat flow in Earth's crust.

\section{Conclusion}

As detailed in this document and elsewhere~\cite{prl}, the KDK collaboration has successfully measured the branching ratio of the elusive electron-capture decay of \K\ to the ground state of \Ar. The measured branching ratio is in agreement with the theoretical value calculated in this work.  When compared to the traditionally used  branching ratio,  which is a factor of two larger, a factor of five increase in precision has been achieved. The improved precision will allow  rare-event searches to better understand their low-energy backgrounds. Additionally, this measurement represents the first experimental verification of a third-forbidden unique transition informing nuclear structure theory. Finally, our measurement  affects K-decay based geochronological estimations by up to a percent.

%




\begin{acknowledgments}

Xavier Mougeot of LNHB drew our attention to his latest evaluation of the decay scheme of \K.
Engineering support has been contributed by Miles Constable and Fabrice R\'eti\`ere of TRIUMF, as well as by Koby Dering through the NSERC/Queen’s MRS.
Funding in Canada has been provided by NSERC through SAPIN and SAP RTI grants, as well as by the Faculty of Arts and Science of Queen's University, and by the McDonald Institute.
Work was performed at Oak Ridge National Laboratory, managed by UT-Battelle, LLC, for the U.S. Department of Energy under Contract DE-AC05-00OR22725. Thermal deposition was conducted at the Center for Nanophase Materials Sciences, which is a DOE Office of Science User Facility. 
This manuscript has been authored by UT-Battelle, LLC under Contract No. DE-AC05-00OR22725 with the U.S. Department of Energy. 
U.S. support has also been supplied by the Joint Institute for Nuclear Physics and Applications, and by NSF grant EAR-2102788.
This material is based upon work supported by the U.S. Department of Homeland Security under grant no. 2014-DN-077-ARI088-01.
J.C., L.E.M., and P.R.R. acknowledge support from NSF grant 2102788.

The United States Government retains and the publisher, by accepting the article for publication, acknowledges that the United States Government retains a non-exclusive, paid-up, irrevocable, world-wide license to publish or reproduce the published form of this manuscript, or allow others to do so, for United States Government purposes. The Department of Energy will provide public access to these results of federally sponsored research in accordance with the DOE Public Access Plan~\cite{DOE_PublicAccess}.
The views and conclusions contained in this document are those of the authors and should not be interpreted as necessarily representing the official policies, either expressed or implied, of the U.S. Department of Homeland Security.
Any use of trade, firm, or product names is for descriptive purposes only and does not imply endorsement by the U.S. Government.

\end{acknowledgments}

\bibliography{bibliography}

\end{document}